\documentclass[aps,prd,groupedaddress, showkeys, showpacs]{revtex4}
\usepackage{grffile,graphicx,subfigure,enumerate,pstricks,latexsym,longtable}
\usepackage[english]{babel}
\usepackage{epstopdf}
\usepackage{amsmath,amssymb}

\def\beq{\begin{equation}}
\def\eeq{\end{equation}}
\def\bea{\begin{eqnarray}}
\def\eea{\end{eqnarray}}

\begin{document}

\title{Constraints on tidal charge of the supermassive black hole at the Galactic Center with trajectories of bright stars}
\author{Alexander F. Zakharov $^{1,2,3,4,5}$}
\email{zakharov@itep.ru}

%\affiliation{$^{1}$ Joint Institute for Nuclear Research, Dubna, Russia}
%\affiliation{$^{2}$ Institute of Physics, Faculty of Philosophy and Science, Silesian University in Opava, Czech Republic}
\affiliation{$^{1}$  National Astronomical Observatories of Chinese Academy of Sciences,
          20A Datun Road,  100012 Beijing, China}
\affiliation{$^{2}$  Institute of Theoretical and Experimental Physics, B. Cheremushkinskaya, 25, 117218, Moscow, Russia}
\affiliation{$^{3}$National Research Nuclear University MEPhI (Moscow Engineering Physics Institute), Kashirskoe highway 31, Moscow, 115409, Russia}
\affiliation{$^{4}$ Joint Institute for Nuclear Research, Dubna, Russia}
\affiliation{$^{5}$North Carolina Central
University, Durham, NC 27707,
 USA}
\date{\today}
\begin{abstract}
As it was pointed out recently in \cite{Hees_PRL_17},   observations of  stars near the Galactic Center with current and future facilities provide an unique tool to test general relativity (GR) and alternative theories of gravity in a strong gravitational field regime.
In particular, the authors showed that the Yukawa gravity could be constrained with Keck and TMT observations.
Some time ago, Dadhich et al. showed in \cite{Dadhich_01} that  the Reissner -- Nordstr\"om metric with a tidal charge is naturally appeared in the framework of Randall -- Sundrum model with an extra dimension ($Q^2$ is called tidal charge and it could be negative in such an approach).
Astrophysical consequences of presence of black holes with a tidal charge are considerered, in particular, geodesics and shadows in Kerr -- Newman braneworld metric
are analyzed in \cite{Schee_09_IJMPD}, while profiles of emission lines generated by rings orbiting braneworld Kerr black hole are considered in \cite{Schee_09_GRG}.
Possible observational signatures of gravitational lensing in a presence of the Reissner -- Nordstr\"om black hole with a tidal charge at the Galactic Center are discussed in papers  \cite{Bin-Nun_10,Bin-Nun_10_2,Bin-Nun_11}. Here we are following such an approach and we obtain analytical expressions for orbital precession for Reissner -- Nordstr\"om -- de-Sitter solution  in post-Newtonian approximation
  and discuss opportunities to constrain parameters of the metric from observations of bright stars with current and future astrometric observational facilities such as VLT, Keck, GRAVITY, E-ELT and TMT.

\end{abstract}
%overview

\pacs{04.70.Bw, 04.25.-g, 04.70.-s, 97.60.Lf \hfill
}
\maketitle
%\tableofcontents

\section{Introduction}

The Galactic Center is a very peculiar object.
A couple of different  models  have been suggested for it, including dense cluster of stars \cite{Reid_09}, fermion ball
\cite{Munyaneza_02}, boson stars \cite{Jetzer_92,Torres_00}, neutrino balls \cite{DePaolis_01}. Later, some of these models have been constrained with subsequent observations \cite{Reid_09}. However, as it was found in computer simulations, sometimes differences for alternative models may be very tiny
as it was shown in paper \cite{Vincent_16} where the authors discussed shadows for boson star and black hole models.
 The most natural and generally accepted model for the Galactic Center  is a supermassive black hole  (see, e.g. recent reviews \cite{Goddi_17,Eckart-etal:2017:FOUNPH:,Zakharov_MIFI_17,Zakharov_MIFI_18}).
A natural way to evaluate a gravitational potential is to analyze trajectories of photons or test particles   moving in the potential.
Shapes of shadows forming by photons moving around black holes were discussed in   \cite{Bardeen_73,Chandrasekhar_83,Falcke_00,Melia_01} (see also \cite{ZNDI_05}). Shadows (dark spots) can not be detected but theoretical models could describe a distribution of bright structures around these dark shadows.
Bright structures around shadows  are being observing with an improving accuracy of current and forthcoming VLBI facilities in mm-band, including the Event Horizon Telescope
\cite{Doeleman_08,Doeleman_08b,Doeleman_09,Doeleman_17}.

To create an adequate theoretical model for the Galactic Center astronomers monitored  trajectories of bright stars (or clouds of hot gas) using the largest telescopes VLT and Keck with adaptive optics facilities \citep{Ghez_00,Ghez_03,Ghez_04,Ghez_05,Weinberg_05,Meyer_12,Morris_12}.
One could introduce a distance between observational data for trajectories of bright stars and their theoretical models.
Practically, such a distance is a measure of quality for a theoretical fit.
To test different theoretical models one of the most
simple approach is to compare apocenter (pericenter) shifts for theoretical fits and observational data for trajectories.
If an apocenter (pericenter) shifts for a theoretical fit exceed   apocenter (pericenter) shifts obtained from observations
one should rule out these interval for parameters for theoretical fits.
Based on such an approach one could evaluate parameters of black hole, stellar cluster and dark matter cloud around the Galactic Center  because if there is an extended mass distribution inside a bright star orbit in addition to black hole, the extended mass distribution causes an apocenter shift in direction which is opposite to relativistic one \cite{ZNDI_PRD_2007,NDIQZ_07}.
One could also check predictions of general relativity or alternative theories of gravity. For instance, one could evaluate
constraints on parameters of $R^n$ theory, Yukawa gravity and graviton masses with trajectories of bright stars at the Galactic Center because in the case of alternative theories of gravity a weak gravitational field limit differs from Newtonian one, so
trajectories of bright stars differ from elliptical ones and analyzing observational data with theoretical fits obtained in the framework of alternative theories of gravity one constrains parameters of such theories \cite{BJBZ_PRD_12,BJBJZ_JCAP_13,ZBBJJ_ASR_14,ZJBBJ_JCAP_16,Zakharov_Quarks_16,Zakharov_JCAP_18}
(see, also  discussion of observational ways to investigate opportunities to find possible deviations from general relativity with observations of bright stars at the Galactic Center \cite{Hees_PRL_17,Hees_17}).

In paper \cite{Dadhich_01} it was shown that  the Reissner -- Nordstr\"om metric with a tidal charge could arise  in Randall -- Sundrum model with an extra dimension.
Braneworld black holes are considered assuming that they could substitute conventional black holes in astronomy, in particular, geodesics and shadows in Kerr -- Newman braneworld metric
are analyzed in \cite{Schee_09_IJMPD}, while profiles of emission lines generated by rings orbiting braneworld Kerr black hole are considered in \cite{Schee_09_GRG}.
 Later it was proposed to consider signatures of gravitational lensing assuming a presence of the Reissner -- Nordstr\"om black hole with a tidal charge at the Galactic Center \cite{Bin-Nun_10,Bin-Nun_10_2,Bin-Nun_11}.
In paper  \cite{Zakharov_PRD_2014} analytical expressions for shadow radius of   Reissner -- Nordstr\"om black hole have been derived while shadow sizes  for Schwarzschild -- de Sitter (K\"ottler) metric have been found in papers \cite{Stuchlik_83,Zakharov_2014}.
In the paper for a particle motion in Reissner -- Nordstr\"om -- de-Sitter metric we derive analytical expressions for orbital precession and discuss constraints on tidal charge from current and future observations of bright stars near the Galactic Center.

%%%%%%%%%%%%%%%%%%%%%%%%%%%%%%%%%%%%%%%%%%%%%%%%%%%%%%%%%%%%%%%%%%%%%%%%%%%
\section{Basic notations} \label{sec1}
%%%%%%%%%%%%%%%%%%%%%%%%%%%%%%%%%%%%%%%%%%%%%%%%%%%%%%%%%%%%%%%%%%%%%%%%%%%

            We use a system of units where $G = c= 1$.
The line element of the spherically symmetric Reissner -- Nordstr\"om -- de-Sitter  metric is
\beq
d s^2 = - f(r) d t^2 + f(r)^{-1} d r^2 + r^2 d\theta^2 + r^2 \sin^2\theta d\phi^2 , \label{metric_RN}
\eeq
where  function $f(r)$ is defined as
\beq
f(r) = 1 - \frac{2M}{r}  + \frac{Q^2}{r^2} - \frac{1}{3} \Lambda r^2. \label{function_f}
\eeq
Here $M$ is a black hole mass, $Q$ is its charge and $\Lambda$ is cosmological constant.
In the case of a tidal charge \cite{Dadhich_01}, $Q^2$ could be negative.
Similarly to \cite{Carter_73,Sharp_79,Stuchlik_83,Stuchlik_02}, geodesics could be obtained the Lagrangian
\begin{equation}
{\cal{L}}=
 -\frac{1}{2} g_{\mu \nu} \frac{d x^\mu}{d \lambda} \frac{d x^\nu}{d \lambda},
 \label{Lagrangian-1}
\end{equation}
where $ g_{\mu \nu}$ are the components of metric (\ref{metric_RN} and $\lambda$ is the affine parameter.
There are three constants of motion for geodesics which come from the  metric (\ref{metric_RN}), namely
\begin{equation}
 g_{\mu \nu} \dfrac{d x^\mu}{d \lambda} \dfrac{dx^\nu}{d \lambda}=-m,
\label{mass-1}
\end{equation}
which is a test particle mass
and two constants connected with an independence of the metric on  $\phi$ and $t$ coordinates, respectively
\begin{equation}
 g_{\phi \nu} \frac{d x^\nu}{d \lambda}=h,
\label{momentum-1}
\end{equation}
and
\begin{equation}
 g_{t \nu} \frac{d x^\nu}{d \lambda}=E.
\label{Energy}
\end{equation}
For vanishing $\Lambda$-term these integrals of motion ($h$ and $E$) could be interpreted as angular momentum and energy of a test particle, respectively.
Geodesics for massive particles could be written in the following form
\begin{equation}
\label{radial_coordinate_equation}
r^4 \left(\dfrac{dr}{d \lambda}\right)^2 = E^2r^4 - \Delta (m^2r^2+h^2),
\end{equation}
where
\begin{equation}
\label{Delta}
\Delta=\left(1-\frac{1}{3}\Lambda r^2\right)r^2 - 2M r+Q^2.
\end{equation}
or we could write Eq. (\ref{radial_coordinate_equation}) in the
following form
\begin{equation}
\label{radial_coordinate_equation2}
r^4 \left( \dfrac{dr}{d \tau}\right)^2 = (\hat{E}^2-1)r^4 +{2Mr^3}- Q^2r^2 +\frac{1}{3} \Lambda r^6 -\hat{h}^2(r^2-\frac{\Lambda}{3}r^4-2Mr+Q^2),
\end{equation}
where $\tau=m \lambda$ is the proper time,  $\hat{E}=\dfrac{E}{m}$ and $\hat{h}=\dfrac{h}{m}$. We will omit symbol $\wedge$ below.
Since
\begin{equation}
\label{radial_coordinate_equation3}
r^4 \left(\dfrac{d \phi}{d \tau}\right)^2 = h^2,
\end{equation}
one could obtain
\begin{equation}
\label{radial_coordinate_equation4}
\left(\dfrac{d r}{d \phi}\right)^2 = \frac{1}{{h}^2}({E}^2-1)r^4 +\frac{2Mr^3}{{h}^2}- \frac{Q^2r^2}{{h}^2} +\frac{1}{3h^2} \Lambda r^6 -(r^2-\frac{\Lambda}{3}r^4-2Mr+Q^2),
\end{equation}
It is convenient to introduce new variable $u=1/r$. Since
\begin{equation}
 \left(\dfrac{d u}{d \lambda}\right)^2 = \left(\dfrac{d r}{d \phi}\right)^2 u^4,
\label{u_variable}
\end{equation}
one obtains
\begin{equation}
 \left(\dfrac{d u}{d \lambda}\right)^2 =
 \frac{1}{h^2}({E}^2-1) +\frac{2Mu}{{h}^2}- \frac{Q^2u^2}{{h}^2} +\frac{\Lambda }{3h^2 u^2}  -(u^2-\frac{\Lambda}{3}-2Mu^3+Q^2u^4),
 % (\hat{E}^2-1)r^4 +{2Mr^3}- Q^2r^2 -\frac{1}{3} \Lambda r^6 -\hat{h}^2(r^2-\frac{\Lambda}{3}r^4-2Mr+Q^2),
\label{u_variable2}
\end{equation}
therefore,
\begin{equation}
 \dfrac{d^2 u}{d \lambda^2} +u = \frac{M}{{h}^2}+ 3Mu^2  - \frac{Q^2u}{{h}^2}  -2Q^2u^3 -\frac{\Lambda }{3h^2 u^3} ,
 % (\hat{E}^2-1)r^4 +{2Mr^3}- Q^2r^2 -\frac{1}{3} \Lambda r^6 -\hat{h}^2(r^2-\frac{\Lambda}{3}r^4-2Mr+Q^2),
\label{u_variable3}
\end{equation}
and as it is noted in \cite{Adkins_07} the first term in the right hand side of Eq. (\ref{u_variable3}) corresponds to the Newtonian case, the second term corresponds to
the GR correction from the Schwarzschild metric (see also book \cite{Anderson_67}), meanwhile one could see inspecting Eq. (\ref{u_variable3}) that  third and forth term correspond to a presence of $Q$ parameter in metric (\ref{metric_RN}),
the fifth term corresponds to a $\Lambda$-term presence in the metric.
Assuming that second, third, forth and fifth terms in the right hand side of Eq. (\ref{u_variable3}) are small in respect to the basic Newtonian solution, one could evaluate relativistic precession
for each term and after that one has to calculate an algebraic sum of all shifts induced by different terms.

\section{Relativistic precession evaluation}

An impact of non-vanishing charge in Reissner -- Nordstr\"om metric on orbital precession was discussed in papers \cite{Burman_69,Teli_84,Rathod_89,Dean_99,Gong_09} considering perturbations of Schwarzschild metric, see for instance  \cite{Anderson_67,Treder_80}.
However, Eq. (\ref{u_variable3}) was not considered in these papers.
When people discussed astrophysical consequences of this effect they evaluated an impact of Solar charge on Mercury  precession orbit \cite{Burman_69} and it is clear that the effect is very small due to constraints on Solar electric charge. Similarly, for astrophysical black holes including the black holes at the Galactic Center, their electric charges are expecting to be vanishing or very small. However, significant tidal charges $|Q|$ which
are comparable with $M$ are discussed in the literature \cite{Bin-Nun_10_2,Bin-Nun_11} where the author discussed an opportunity to
evaluate a tidal charge $Q^2 \approx - 6.4 M^2$ or $Q^2 \approx 1.6 M^2$ from gravitational lensing.

An expression for apocenter (pericenter) shifts for Newtonian potential plus
small perturbing function is given as a solution in the classical (L \& L) textbook
\cite{Landau_76} (see also applications of the expressions for
calculations of stellar orbit precessions in presence of the the supermassive
black hole and dark matter at the Galactic Center \cite{Dokuchaev_15,Dokuchaev_15a}).
In paper \cite{Adkins_07}, the authors derived the expression which
is equivalent to the (L \& L) relation and which can be used for our needs.
According to the procedure proposed in  \cite{Adkins_07}
one could re-write Eq. (\ref{u_variable3}) in the following form
\begin{equation}
 \dfrac{d^2 u}{d \tau^2} +u = \frac{M}{{h}^2}- \frac{g(u)}{{h}^2},
\label{u_variable4}
\end{equation}
where $g(u)$ is a perturbing function which is supposed to be small and it could be presented as a conservative force
in the following form
\begin{equation}
 g(u)=r^2 F(r)|_{r=1/u}, \quad F(r)=-\frac{dV}{dr}.
\label{u_variable5}
\end{equation}
For potential $V(r)=\dfrac{\alpha_{-(n+1)}}{r^{-(n+1)}}$ (where $n$ is a natural number)
one obtains \cite{Adkins_07}
\begin{equation}
 \Delta \theta (-(n+1))=\frac{-\pi \alpha_{-(n+1)}\chi^2_n(e)}{ML^n},
\label{Delta_negative power}
\end{equation}
where
\begin{equation}
 \chi^2_n(e)=n(n+1)_2F_1\left(\frac{1}{2}-\frac{n}{2},\frac{1}{2}-\frac{n}{2},2,e^2\right),
\label{chi_n}
\end{equation}
$_2F_1$ is the Gauss hypergeometrical function, $L$ is the semilatus rectum ($L=h^2/M$) and we have $L=a(1-e^2)$ ($a$ is semi-major axis and $e$ is eccentricity). An alternative approach for evaluation of pericenter advance within of  Rezzolla -- Zhidenko (RZ) parametrization \cite{Rezzolla_14}
has been described in \cite{DeLaurentis_17} for theoretical analysis of pulsar timing in the case if pulsars are moving in the strong gravitational field of the supermassive black hole at the Galactic Center. Since pulsars are very precise and stable clocks, studies of pulsar timing gives an opportunity to investigate gravitational field in the vicinity of the supermassive black hole.

In paper \cite{Adkins_07} the authors  obtained orbital precessions for positive powers of perturbing function
\begin{equation}
 \Delta \theta (n)=\frac{-\pi \alpha_{n}a^{n+1} \sqrt{1-e^2}\chi^2_n(e)}{M}.
\label{Delta_negative power_2}
\end{equation}

For GR term in Eq. (\ref{u_variable3}) the perturbing potential is
$V_{GR}(r)=-\dfrac{Mh^2}{r^3}$
and
one obtains the well-known result $n=2$ (see, for instance \cite{Adkins_07} and  textbooks on GR)
\begin{equation}
  \Delta \theta (GR) := \Delta \theta (-(3))=\frac{6\pi M}{L}.
\label{Delta_GR}
\end{equation}
For the third term in Eq. (\ref{u_variable3}) one has potential $V_{RN1}(r)=\dfrac{Q^2}{2r^2}$ ($\alpha_{-2}=\dfrac{Q^2}{2}$ and $n=1$), therefore, one obtains
\begin{equation}
  \Delta \theta (RN1) := \Delta \theta (-(2))_{RN1}=-\frac{\pi Q^2}{ML}.
\label{Delta_RN1}
\end{equation}
Eq. (\ref{Delta_RN1}) was derived earlier in \cite{Burman_69} to evaluate an impact of Solar charge on orbital precession of Mercury, however, we re-derive the Eq. (\ref{Delta_RN1}) following a procedure suggested in \cite{Landau_76} since this approach is more clear and it could be applied for
other types of perturbing potentials.
For the forth term in Eq. (\ref{u_variable3}) one has potential $V_{RN2}(r)=\dfrac{h^2Q^2}{2r^4}$  ($\alpha_{-4}=\dfrac{h^2Q^2}{2}$ and $n=3$) , therefore, one obtains
\begin{equation}
  \Delta \theta (RN2) := \Delta \theta (-(4))_{RN2}=-\frac{3\pi Q^2 (4+e^2)}{2L^2}.
\label{Delta_RN2}
\end{equation}
Since according to our assumptions $M \ll L$, one has $\dfrac{Q^2}{L^2} \ll \dfrac{Q^2}{ML}  $ and we ignore the apocenter (pericenter) shift which is described with Eq. (\ref{Delta_RN2}).
For the fifth (de-Sitter or anti-de-Sitter) term in Eq. (\ref{u_variable3}) one has potential $V_{dS}(r)=-\dfrac{\Lambda r^2}{6}$ ($\alpha_2=-\dfrac{\Lambda}{6}$) and
one has the corresponding apocenter (pericenter) shift
\begin{equation}
  \Delta \theta (\Lambda) := \Delta \theta (2)_{dS}=\frac{\pi \Lambda a^3 \sqrt{1-e^2}}{M}.
\label{Delta_dS}
\end{equation}
Eq. (\ref{Delta_dS}) was derived earlier in \cite{Kerr_03} and re-derived in  \cite{Adkins_07} with (L \& L) approach \cite{Landau_76}.
In paper \cite{Sereno_06} Eq. (\ref{Delta_dS}) was used to discuss consequences  of a non-vanishing $\Lambda$-term from observations in Solar system.

Therefore, a total shift of a pericenter is
\begin{equation}
  \Delta \theta (total) :=   \frac{6\pi M}{L}                 -\frac{\pi Q^2}{ML}         +   \frac{\pi \Lambda a^3 \sqrt{1-e^2}}{M}.
\label{Delta_dS_2}
\end{equation}
and one has a relativistic advance for a tidal charge with $Q^2 < 0$ and apocenter shift dependences on eccentricity and semi-major axis
are the same for GR and Reissner -- Nordstr\"om advance but corresponding factors (${6\pi M}$  and    $-\dfrac{\pi Q^2}{M}$) are different, therefore,
it is very hard to distinguish a presence of a tidal charge and black hole mass evaluation uncertainties.
For $Q^2 > 0$, there is an apocenter shift in the opposite direction in respect to GR advance.
As it was noted each term in Eq. (\ref{Delta_dS_2}) was known earlier, but people did not consider them together perhaps because of  small values electric charge and $\Lambda$-term. However,  a wider range for tidal charge was considered for the black hole at the Galactic Center \cite{Bin-Nun_10_2,Bin-Nun_11} and an excellent precision of astrometrical observations has been reached in last years and it gives an opportunity
to evaluate parameters of alternative theories of gravity with these observations.

\section{Estimates}

As it was noted by the astronomers of the Keck group \cite{Hees_PRL_17}, pericenter shift has not be found yet for S2 star, however,
an upper confidence limit on a linear drift is constrained
\begin{equation}
  | \dot{\omega}| < 1.7 \times 10^{-3} {\rm rad}/{\rm yr}.
\label{observational_constraint}
\end{equation}
at 95\% C.L., while GR advance for the pericenter
is  \cite{Hees_17}
\begin{equation}
  | \dot{\omega}_{GR}| = \frac{6\pi GM}{Pc^2(1-e^2)}=1.6 \times 10^{-4} {\rm rad}/{\rm yr},
\label{GR_constraint}
\end{equation}
where $P$ is the orbital period for S2 star (in this section we use dimensional constants $G$ and $c$ instead of geometrical units).
Based on such estimates one could constrain alternative theories of gravity following the approach used in \cite{Hees_PRL_17}.

\subsection{Estimates of (tidal) charge constraints}

Assuming  $\Lambda=0$ we consider constraints on $Q^2$ parameter from previous and future observations of S2 star.
One could re-write orbital precession in dimensional form
\begin{equation}
   \dot{\omega}_{RN} = \frac{\pi Q^2}{PGML},
\label{RN_constraint_0}
\end{equation}
where $P$ is an orbital period.
Taking into account a sign of pericenter shift
for a tidal charge with $Q^2 < 0$, one has
\begin{equation}
   \dot{\omega}_{RN} < 1.54 \times 10^{-3} {\rm rad}/{\rm yr} \approx 9.625~ \dot{\omega}_{GR},
\label{TRN_constraint_1}
\end{equation}
therefore,
\begin{equation}
    -57.75 M^2 < Q^2 <0,
\label{TRN_constraint}
\end{equation}
with $ 95 \%$ C. L.
For $Q^2 >0$, one has
\begin{equation}
  | \dot{\omega}_{RN}| < 1.86 \times 10^{-3} {\rm rad}/{\rm yr} \approx 11.625~ \dot{\omega}_{GR},
\label{RN_constraint_2}
\end{equation}
therefore,
\begin{equation}
    0 < |Q| < 8.3516 M,
\label{RN_constraint}
\end{equation}
with $ 95 \%$ C. L.
As it was noted in \cite{Hees_PRL_17} in 2018 after the pericenter passage of S2 star the current uncertainties of
$| \dot{\omega}|$ will be improved by a factor 2, so
for a tidal charge with $Q^2 < 0$, one has
\begin{equation}
   \dot{\omega}_{RN} < 6.9 \times 10^{-4} {\rm rad}/{\rm yr} \approx 4.31~ \dot{\omega}_{GR},
\label{TRN_constraint_1_18}
\end{equation}
\begin{equation}
    -25.875 M^2 < Q^2 <0,
\label{TRN_constraint_18}
\end{equation}
For $Q^2 >0$, one has
\begin{equation}
  | \dot{\omega}_{RN}| < 9.1 \times 10^{-4} {\rm rad}/{\rm yr} \approx 5.69~ \dot{\omega}_{GR},
\label{RN_constraint_2_18}
\end{equation}
therefore,
\begin{equation}
    0 < |Q| < 5.80 M,
\label{RN_constraint_18}
\end{equation}
One could expect that subsequent observations with VLT, Keck, GRAVITY, E-ELT and TMT will
significantly improve an observational constraint on
$ | \dot{\omega}|$, therefore, one could expect that a range of possible values of $Q$ parameter would be essentially reduced.

As it was noted in paper \cite{Hees_PRL_17},  currently Keck astrometric uncertainty is around $\sigma = 0.16$~mas, therefore,
an angle $\delta =2 \sigma$ (or two standard deviations)  is measurable  with  around 95\% C.L.
 In this case $\Delta \theta (GR)_{S2}= 2.59 \delta $ for S2 star  where we adopt $\Delta \theta (GR)_{S2} \approx 0.83$. Assuming that GR predictions about
orbital precession will be confirmed in the next 16 years with $\delta$ accuracy (or $ \left|\dfrac{\pi Q^2}{ML}\right|  \lesssim \delta$), one could constrain  $Q$ parameter
\begin{equation}
    |Q^2| \lesssim 2.32 M^2,
\label{RN_constraint_34}
\end{equation}
where we wrote absolute value of $Q^2$ since for a tidal charge $Q^2$ could be negative. For negative $Q^2$ this estimate is better than estimate considered
in \cite{Bin-Nun_10_2} ($Q^2 \approx - 6.4 M^2$), however,  the estimate (\ref{RN_constraint_34}) is slightly more worse than  $Q^2 \approx 1.6 M^2$.

If we adopt   uncertainty  $\sigma_{TMT} = 0.015$~mas for TMT-like scenario as it was used in  \cite{Hees_PRL_17} ($\delta_{TMT}=2\sigma_{TMT}$)
or in this case $\Delta \theta (GR)_{S2}= 27.67 \delta_{TMT} $ for S2 star and assuming again that GR predictions
about orbital precession of S2 star will be confirmed  with  $\delta_{TMT}$ accuracy (or $ \left|\dfrac{\pi Q^2}{ML}\right|  \lesssim \delta_{TMT}$) , one could conclude that
\begin{equation}
    |Q^2| \lesssim 0.216 M^2,
\label{RN_constraint_TMT}
\end{equation}
or based on results of  future observations one could expect to reduce significantly a possible range of $Q^2$ parameter in comparison with
a possible hypothetical range of $Q^2$ parameter which was discussed in \cite{Bin-Nun_10,Bin-Nun_10_2}.

Recently the GRAVITY team reported about a discovery of post-Newtonian gravitational redshift near S2 star pericenter passage \cite{Gravity_18}.
Assuming $f=0$ corresponds to the Newtonian case and $f=1$ corresponds to the first post-Newtonian correction of GR, the GRAVITY collaboration estimated  $f$-value from observational data comparing precessions for Schwarzschild and Newtonian approaches and they concluded that the $f$-value must be much closer to GR value or more precisely $f=0.94 \pm 0.09$ \cite{Gravity_18} (see also discussions in \cite{Zakharov_Quarks_18}).
If we adopt   uncertainty  $\sigma_{\rm GRAVITY} = 0.030$~mas of the GRAVITY facilities \cite{Gravity_18} and assuming again that GR predictions
 on orbital precession of S2 star will be confirmed  with  $\delta_{\rm GRAVITY}=2\sigma_{\rm GRAVITY}$ accuracy (or $ \left|\dfrac{\pi Q^2}{ML}\right|  \lesssim \delta_{\rm GRAVITY}$), one could conclude that
\begin{equation}
    |Q^2| \lesssim 0.432 M^2,
\label{RN_GRAVITY}
\end{equation}
or based on results of  future GRAVITY observations one could expect to reduce significantly a possible range of $Q^2$ parameter in comparison with
a possible range of $Q^2$ parameter constrained with current and future Keck data.

\subsection{Estimates of $\Lambda$-term constraints}

In this subsection we assume that $Q=0$.
One could re-write orbital precession in dimensional form
\begin{equation}
   \dot{\omega}_{\Lambda} = \frac{\pi \Lambda c^2 a^3 \sqrt{1-e^2}}{PGM},
\label{DE_constraint}
\end{equation}
Dependences  of functions  $\dot{\omega_{\Lambda}}$ and  $\dot{\omega_{GR}}$ on eccentricity and semi-major axis are different and orbits
with higher semi-major axis and smaller eccentricity could provide a better estimate of  $\Lambda$-term (the S2 star orbit has a rather high eccentricity). However, we use observational constraints
for $S2$ star. For positive $\Lambda$, one has relativistic advance and
\begin{equation}
   \dot{\omega_{\Lambda}} < 1.54 \times 10^{-3} {\rm rad}/{\rm yr} \approx 9.625~ \dot{\omega_{GR}},
\label{DE_constraint_1}
\end{equation}
or
\begin{equation}
 0 < {\Lambda} < 3.9 \times 10^{-39} {\rm cm}^{-2},
\label{DE_constraint_2}
\end{equation}
for $\Lambda <0$ one has
\begin{equation}
 0 < -\Lambda < 4.68 \times 10^{-39} {\rm cm}^{-2},
\label{DE_constraint_3}
\end{equation}

if we use current accuracy of Keck astrometric measurements $\sigma = 0.16$~mas and monitor S2 star for 16 years
and assume that additional apocenter shift ($2\sigma$)could be caused by a presence of $\Lambda$-term, one obtains
\begin{equation}
 |{\Lambda}| < 1.56 \times 10^{-40} {\rm cm}^{-2},
\label{DE_constraint_4}
\end{equation}
while for TMT-like accuracy $\delta_{TMT} = 0.015$~mas
one has
\begin{equation}
 |{\Lambda}| < 1.46 \times 10^{-41} {\rm cm}^{-2}.
\label{DE_constraint_5}
\end{equation}
As one can see, constraints on cosmological constant from orbital precession of bright stars near the Galactic Center are much weaker
than not only its cosmological estimates but also than its estimates from Solar system data \cite{Sereno_06}.

%%%%%%%%%%%%%%%%%%%%%%%%%%%%%%%%%%%%%%%%%%%%%%%%%%%%%%%%%%%%%%%%%%%%%%%%%%%
\section{Conclusions} \label{Conclusion}
%%%%%%%%%%%%%%%%%%%%%%%%%%%%%%%%%%%%%%%%%%%%%%%%%%%%%%%%%%%%%%%%%%%%%%%%%%%

We consider the first relativistic corrections for apocenter shifts in post-Newtonian approximation for the case of Reissner -- Nordstr\"om -- de-Sitter metric.
Among different theoretical models have been proposed for the Galactic Center
different black hole models are rather natural. Perhaps, assumptions
about a presence of electric charge in the metric do not look very realistic
because  a space media is usually quasi-neutral, but the charged black holes are discussed in the literature see, for instance \cite{Moradi_17} and references therein.
Moreover, a Reissner -- Nordstr\"om metric could arise in a natural way in alternative theories of gravity like Reissner -- Nordstr\"om
solutions with a tidal charge in Randall--Sundrum model  \cite{Dadhich_01} (such an approach is widely discussed in the literature).
Recently, it was found that Reissner -- Nordstr\"om metric is a rather natural solution in Horndeski gravity
\cite{Babichev_17} and in this case
$Q^2$ parameter reflects an interaction with a scalar field and it could be also negative similarly to a tidal charge.
In paper \cite{Babichev_17} it was expressed an opinion that the hairy black hole solutions look rather realistic and
these objects could exist in centers of galaxies and if such objects (hairy black holes in Horndeski gravity) exist in nature, in particular in the Galactic Center, current and future advanced facilities such as GRAVITY \cite{blin15}, E-ELT \cite{eelt14}, TMT \cite{tmt14} etc.
may be very useful to detect signatures of black hole hairs of an additional dimension. Therefore, non-vanishing (positive or negative) $Q^2$ parameter is arisen due to a presence of extra dimension or in Horndeski gravity for black holes with a scalar hair. We outline a procedure to constrain
$Q^2$ parameter with current and future observations of bright stars at the Galactic Center. Even current Keck facilities could constrain
$Q^2$ better ($Q^2 \approx -2.32 M^2$) than with analysis of hypothetical variations of S2 brightness as it was suggested in \cite{Bin-Nun_10_2}.

Certainly, $\Lambda$-term
should be present in the model, however, if we adopt its cosmological value it should be very tiny
to cause a significant impact on relativistic precession for trajectories of bright stars.
If we have a dark energy instead of cosmological constant, one should
propose ways to evaluate dark energy for different cases, therefore, one could constrain $\Lambda$-term
from observations as it was noted in \cite{Zakharov_2014} analyzing impact of $\Lambda$-term on observational phenomena
near the Galactic Center (similarly to the cases where an impact of $\Lambda$-term has been analyzed for effects in Solar system \cite{Kagramanova_06,Jetzer_06,Sereno_06}).

%%%%%%%%%%%%%%%%%%%%%%%%%%%%%%%%%%%%%%%%%%%%%%%%%%%%%%%%%%%%%%%%%%%%%%%%%%%
\section*{Acknowledgments}
%%%%%%%%%%%%%%%%%%%%%%%%%%%%%%%%%%%%%%%%%%%%%%%%%%%%%%%%%%%%%%%%%%%%%%%%%%%
The author thanks D. Borka, V. Borka Jovanovi\'c,  V. I. Dokuchaev, P. Jovanovi\'c and Z. Stuchl\'ik for useful discussions. A. F. Z. thanks
PIFI grant 2017VMA0014 of Chinese Academy of Sciences at NAOC (Beijing),
NSF
(HRD-0833184) and NASA (NNX09AV07A) at NASA CADRE and NSF CREST
Centers (NCCU, Durham, NC, USA) for a partial support.

The author thanks also an anonymous referee for useful critical remarks.
%%%%%%%%%%%%%%%%%%%%%%%%%%%%%%%%%%%%%%%%%%%%%%%%%%%%%%%%%%%%%%%%%

%%%%%%%%%%%%%%%%%%%%%%%%%%%%%%%%%%%%%%%%%% END %%%%%%%%%%%%%%%%%%%%%%%%%%%%%%%%%%%%%%%

\end{document}